\documentclass[journal,12pt,onecolumn,draftclsnofoot]{IEEEtran}
\IEEEoverridecommandlockouts
\usepackage{cite}
\usepackage{amsmath,amssymb,amsfonts}
\usepackage{algorithmic}
\usepackage{graphicx}
\usepackage{subcaption}
\usepackage{url}
\usepackage{bbm}
\usepackage{bm}
\usepackage{siunitx}
\usepackage{mathtools, nccmath}
\graphicspath{ {Figures/EPS/} }
\usepackage{textcomp}
\usepackage{xcolor}
\usepackage{mathtools}
\usepackage{blindtext}
\usepackage[utf8]{inputenc}
\usepackage{float}
\usepackage{enumitem}
\usepackage{hyperref}
\usepackage{graphicx}
\usepackage{multirow}
\usepackage{tabularx}

\usepackage{acronym}
\acrodef{cdf}[CDF]{cumulative distribution function}
\acrodef{pdf}[PDF]{probability density function}
\acrodef{los}[LOS]{line-of-sight}
\acrodef{nlos}[NLOS]{non-line-of-sight}
\acrodef{ppp}[PPP]{Poisson point process}
\acrodef{bs}[BS]{base station}
\acrodef{mmw}[mmWave]{millimeter waves}
\acrodef{rv}[r.v.]{random variable}
\acrodef{ack}[ACK]{acknowledgement}

\def\BibTeX{{\rm B\kern-.05em{\sc i\kern-.025em b}\kern-.08em
    T\kern-.1667em\lower.7ex\hbox{E}\kern-.125emX}}
\begin{document}
\title{Analysis of Blocking in mmWave Cellular
Systems: Characterization of the LOS and NLOS Intervals in Urban Scenarios\\
\thanks{
Affiliation: Dept. Signal Theory and Communic., Universitat Politècnica de Catalunya, Barcelona, Spain. Emails: \IEEEauthorrefmark{1}cristian.garcia.ruiz@estudiant.upc.edu, \IEEEauthorrefmark{2}antonio.pascual@upc.edu, \IEEEauthorrefmark{3}olga.munoz@upc.edu.

This work has been funded through the projects 5G\&B-RUNNER-UPC (Agencia Estatal de Investigación and Fondo Europeo de Desarrollo Regional, TEC2016-77148-C2-1-R / AEI/FEDER, UE) and ROUTE56 (Agencia Estatal de Investigación, PID2019-104945GB-I00 / AEI / 10.13039/501100011033); and the grant 2017 SGR 578 (AGAUR, Generalitat de Catalunya).

\copyright 2020 IEEE. Personal use of this material is permitted. Permission from IEEE must be obtained for all other uses, in any current or future media, including reprinting/republishing this material for advertising or promotional purposes, creating new collective works, for resale or redistribution to servers or lists, or reuse of any copyrighted component of this work in other works.

DOI: 10.1109/TVT.2020.3037125

}
}

\author{\IEEEauthorblockN{Cristian García Ruiz\IEEEauthorrefmark{1},
Antonio Pascual-Iserte\IEEEauthorrefmark{2}, \IEEEmembership{Senior Member, IEEE},
Olga Muñoz-Medina\IEEEauthorrefmark{3}, \IEEEmembership{Member, IEEE}}}


\markboth{Accepted Paper at IEEE Transactions on Vehicular Technology (Volume: 69, Issue: 12, December 2020)}
{}

\maketitle

\begin{abstract}
In the millimeter waves (mmWave) bands considered for 5G and beyond, the use of very high frequencies results in the interruption of communication whenever there is no line of sight between the transmitter and the receiver. Blockages have been modeled in the literature so far using tools such as stochastic geometry and random shape theory. Using these tools, in this paper, we characterize the lengths of the segments in line-of-sight (LOS) and in non-line-of-sight (NLOS) statistically in an urban scenario where buildings (with random positions, lengths, and heights) are deployed in parallel directions configuring streets.
\end{abstract}


\section{Introduction}
\subsection{Background and motivation}
\IEEEPARstart{W}{ith} the turn of the new decade, we are experiencing a major revolution: the introduction of 5G mobile communications, which is expected to change our daily lives. It will enable new applications such as autonomous driving or remote surgery and will support, also, a high amount of connected devices \cite{modelingmmwave,5G_NR}. Because of this, \ac{mmw} bands above \SI{6}{GHz} are essential to allow for much larger bandwidths and data rates \cite{mmwaveMIMO}. However, these bands entail some drawbacks such as high attenuation, harmful diffraction, and vulnerability to blockages \cite{modelingmmwave}, which occur at those positions where the user and the \ac{bs} or serving access point are in \ac{nlos}.

Previous works in the literature, such as \cite{randomshape,blockageeffects,Article_correlation,relay_positioning}, have studied the probability of being in \ac{los}/\ac{nlos}, modeling the blocking elements, such as buildings, as random objects through stochastic geometry and random shape theory. In particular, they consider that the positions of the blocking elements follow a \ac{ppp} and use random shape theory to model their sizes, shapes, and orientations.

This paper contributes to this topic by statistically characterizing the lengths of the \ac{los} and \ac{nlos} intervals within a linear trajectory in an urban scenario, an aspect not addressed in \cite{randomshape,blockageeffects,Article_correlation,relay_positioning}. Note that, in scenarios where terminals are moving, the lengths of these intervals have an impact on the duration of the time intervals in which the receiver can (or cannot) correctly detect the received signals.

Some related works are \cite{mobileblockers1,mobileblockers2}. In \cite{mobileblockers1}, the authors consider static users and moving punctual blocking elements. Each blocker produces an NLOS period modeled as an exponential \ac{rv} with known mean and, therefore, consolidated tools from queue theory can be employed. In \cite{mobileblockers2}, the authors model mobile blockers as cylinders with given sizes and assume that the statistics of the NLOS time produced by each blocker are also known and given.

In contrast to \cite{mobileblockers1,mobileblockers2}, the goal of this paper is to obtain a justified statistical characterization of the lengths of the \ac{los} and \ac{nlos} intervals in urban environments without any a priori assumption about their distributions or mean values. To that end, and following \cite{randomshape,blockageeffects,Article_correlation,relay_positioning}, we use stochastic geometry and random shape theory to model the buildings that may block the signal transmission. In the study, we assume that the user's trajectory and the buildings are parallel, as happens in urban environments and matching the case of most cities quite well.

\subsection{Goals and contributions}
The contributions of this work are the following derivations:
\begin{itemize}

\item An upper bound of the \ac{cdf} of the length of the LOS intervals derived from the statistics of the sizes and positions of the buildings.
\item An approximation of the average length of the \ac{los} and \ac{nlos} intervals.
\item An approximation of the linear density of the \ac{los} and \ac{nlos} intervals within a trajectory.

\end{itemize}

\subsection{Organization}
Section \ref{sec:Scenario} presents the scenario. Section \ref{sec:prob_segm_los} derives the probability that a segment is in \ac{los}. Using the previous result, an upper bound for the \ac{cdf} of the length of the \ac{los} intervals is obtained in Section \ref{sec:Light}. Section \ref{sec:Average} derives expressions for the average length and density of the \ac{nlos} and \ac{los} intervals. Finally, Section \ref{sec:Results} evaluates the accuracy of the analytic expressions and Section \ref{sec:Conclusions} presents the conclusions.

\subsection{Notation}

\begin{itemize}
	\item $A_S$ denotes the area of a given region $S \subset \mathbb{R}^2$.

\item $K \sim \mathcal{P}\left(\overline{K}\right)$ means that $K$ is a Poisson \ac{rv} with average value $\overline{K}=\mathbb{E}[K]$.

\item $\mathbb{P}(\text{\ac{los}}^{(x)})$ and $\mathbb{P}(\text{\ac{nlos}}^{(x)})$ stand for the probabilities that a point $x$ is in \ac{los} and \ac{nlos}, respectively.


\item A \ac{los} interval is a segment with all its points in \ac{los}. 

\item A NLOS (or blocked) interval is a segment with all its points in \ac{nlos}. 

\end{itemize}

\section{Scenario Description and Blockages Model} \label{sec:Scenario}
\begin{figure}[tbp]
  \centering
  \includegraphics[width=0.73\linewidth]{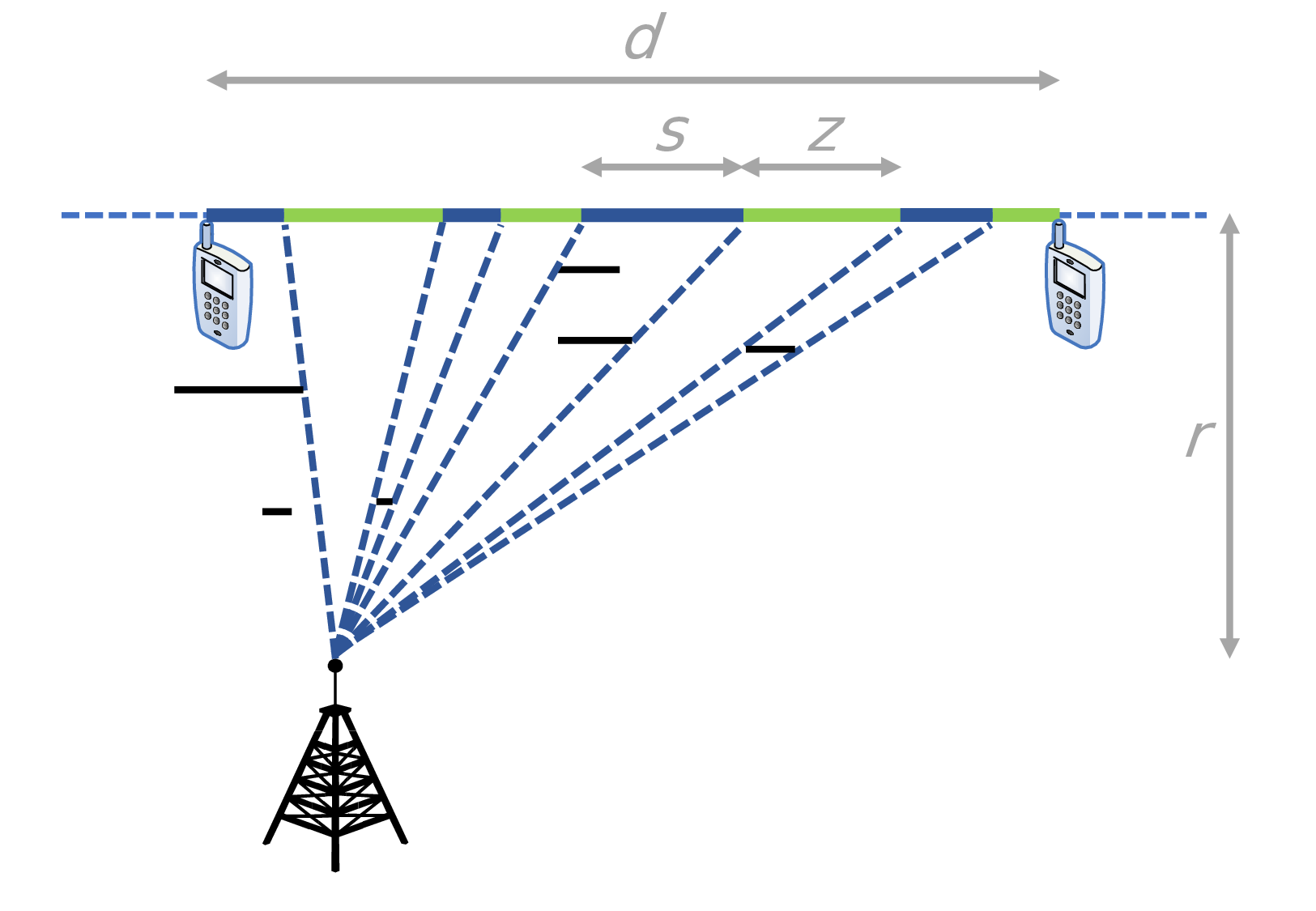}
  \label{fig:Shadowed_trajectory_example_b}
\caption{The buildings (represented in this figure as black lines) block some parts of a trajectory at distance $r$ from the \ac{bs}. The positions and sizes of the blocking elements are random.}
\label{fig:Shadowed_trajectory_example}
\end{figure}

In this paper, we model an urban scenario considering that the positions of the buildings are random and follow a \ac{ppp} distribution with uniform spatial density $\lambda$ (buildings/$m^2$). This implies that the number of buildings in a given region $S$ is a Poisson \ac{rv} with a mean value equal to $\lambda A_S$
\cite{randomshape,blockageeffects,Article_correlation,relay_positioning}.

Modeling the blocking elements as lines with random lengths and heights using random shape theory to obtain the blockage probability at specific positions is a common practice in the related literature \cite{randomshape,blockageeffects,Article_correlation,relay_positioning}. In this paper, we assume that lengths and heights are independent and follow a \ac{pdf} $f_L(l)$ and $f_{H}(h)$, respectively. Regarding their orientation, some papers also consider it random. This assumption may facilitate, in some cases, the computation of the blocking probability; however, it is not the case of the analysis in this paper. Besides, in urban scenarios, buildings and streets are mostly parallel. Therefore, we will consider a parallel orientation for the trajectory and the buildings.

Fig. \ref{fig:Shadowed_trajectory_example} shows an illustrative scenario with linear random blocking elements and a trajectory of length $d$ at minimum distance $r$ from the \ac{bs}. For simplicity, the figure does not consider the heights of the buildings, the \ac{bs}, and the user. Green segments represent segments in \ac{los} (with lengths $z$), whereas blue segments (with lengths $s$) represent segments in \ac{nlos}. Note that the \ac{nlos} segments can correspond to the combined effect of several blocking elements, making the statistical characterization of their lengths more challenging.

\begin{figure}[tbp]
\centerline{\includegraphics[width=10cm]{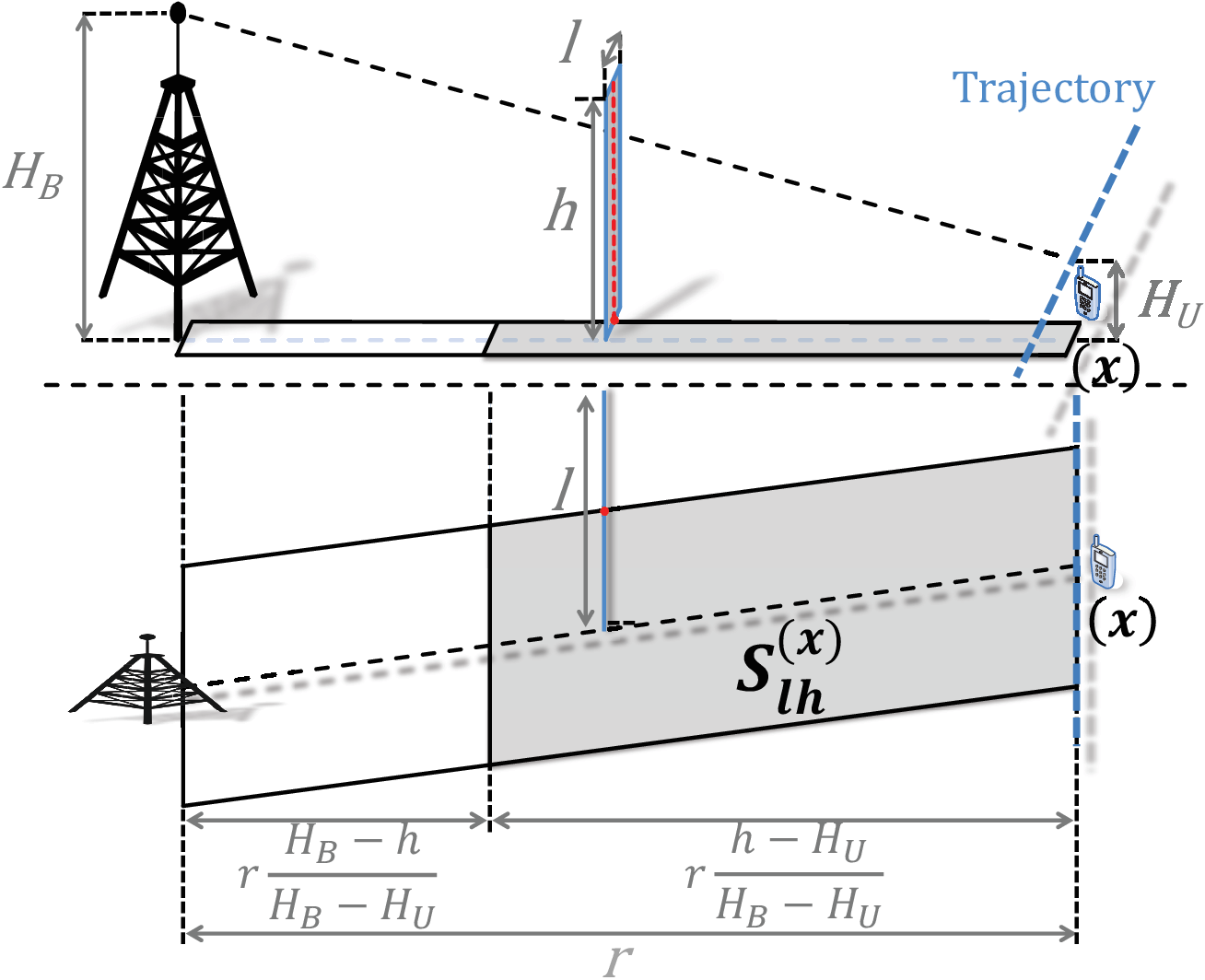}}
\caption{Front view (above) and top view (below) of the blocking region $S_{lh}^{(x)}$, defined as the geometric locus of the centers of the elements of length $l$ and height $h$ able to block the link between the BS and a user located at position $x$.}
\label{fig:Blocking_region_single_point}
\end{figure}

In a real scenario, not only the buildings, but also the \ac{bs} and the users, have a height, denoted here by $H_B$ and $H_U$, respectively. Then, a building between the \ac{bs} and the user will not produce a blocking whenever the building's height is below the vision line between them at the building location, as graphically represented at the upper plot in Fig. \ref{fig:Blocking_region_single_point}. This issue further complicates the statistical characterization of the lengths of the \ac{los} and \ac{nlos} segments in the trajectory.

\section{Probability that a Given Segment Is in LOS} \label{sec:prob_segm_los}
In this section, we compute the probability that a given segment in the trajectory is in \ac{los}. We will first obtain the probability that a given point $x$ in the segment is in \ac{los}. We start by considering blocking elements (that is, buildings) with length $l$ and height $h$, as shown in Fig. \ref{fig:Blocking_region_single_point}. In order not to have a blockage at the point $x$ caused by elements with that shape, no centers of this specific type of blocking elements should fall within the region (a parallelogram) denoted by $S_{lh}^{(x)}$ and represented in grey color in Fig. \ref{fig:Blocking_region_single_point} \cite{randomshape,blockageeffects}.

Let $K_{lh}^{(x)}$ be the number of buildings with lengths and heights in the differential intervals $[l,l+dl]$ and $[h,h+dh]$, respectively, that block point $x$. $K_{lh}^{(x)}$ follows a Poisson distribution with a mean value given by $\mathbb{E}[K_{lh}^{(x)}]=\lambda_{lh}A_{S_{lh}^{(x)}}$ where $\lambda_{lh}=\lambda f_L(l)dl f_{H}(h)dh$ \cite{randomshape}. In the forthcoming analysis, we will consider that the heights of the buildings follow a uniform distribution given by $H \sim \mathcal{U}[H_{\min},H_{\max}]$ and, without loss of generality, $H_U \leq H_{\min}$, $H_U \leq H_B$ and $H_{\min} \leq H_{\max}$.

In general, for a point $x$ to be in \ac{los}, no center of blocking elements can fall within the blocking regions $S_{lh}^{(x)}$ for each possible $l$ and $h$ \cite{randomshape}. Let $K^{(x)}$ denote the number of elements with any length and height that block a point $x$. $K^{(x)}$ also follows a Poisson distribution with mean value \cite{randomshape,blockageeffects}:
\begin{equation}
		\mathbb{E}[K^{(x)}] =  \int_l\int_h\mathbb{E}[K_{lh}^{(x)}]=\lambda\int\int A_{S_{lh}^{(x)}} f_L(l) f_{H}(h) dl dh.\label{eq:meanKx}
\end{equation}

For a trajectory parallel to the orientation of the blocking objects, as usually happens when moving across a city, all the points in the trajectory have the same probability of being in \ac{los}, $\mathbb{P}(\text{\ac{los}}^{(x)})$, and of being in \ac{nlos}, $\mathbb{P}(\text{\ac{nlos}}^{(x)})$. All the points have equal probabilities because, for a given length $l$ and height $h$, the areas of the blocking regions, given by
\begin{equation}\label{eq:A_S_lhtheta_(x)}
	A_{S_{lh}^{(x)}}= 
		\left\{ 
			\begin{array}{lcc}
             			\frac{h-H_U}{H_B-H_U}r \cdot l &\text{if}& H_U \leq h \leq H_B, \\
             			r \cdot l &\text{if}& H_B < h,
             		\end{array}
   		\right.
\end{equation}
are the same for all the points in the segment (i.e., they depend only on $r$, but not on the position of $x$ within the trajectory). According to the previous expressions, (\ref{eq:meanKx}) becomes:
\begin{equation}
	\mathbb{E}[K^{(x)}]= \eta^{(x)}\mathbb{E}[L]r \label{eq:exp_k_(x)}
\end{equation}
with $\eta^{(x)}$ being (see the Appendix for a complete proof)
\begin{equation}\label{eq:eta_x}
	\eta^{(x)}= 
		\left\{
					\begin{array}{l}
             			1 \hspace{4.3cm} \text{if} \hspace{0.3cm} H_B < H_{\min}, \\
             			\frac{2H_BH_{\max}- H_B^2 - H_{\min}^2 - 2H_U(H_{\max}-H_{\min})}{2\left(H_B-H_U \right)\left(H_{\max}-H_{\min}\right)} \\ \hspace{3.2cm} \text{if} \hspace{0.3cm} H_{\min} \leq H_B \leq H_{\max}, \\
             			\frac{H_{\max}+H_{\min}-2H_U}{2\left(H_B-H_U \right)} \hspace{2cm} \text{if} \hspace{0.3cm} H_{\max} < H_B.
             	\end{array}
   		\right.
\end{equation}
Accordingly, for any point $x$ in the given segment, we have
\begin{eqnarray}
 \mathbb{P}(\text{\ac{los}}^{(x)}) &=& \mathbb{P}(K^{(x)}=0)=e^{-\lambda\eta^{(x)}\mathbb{E}[L]r}, \\
	\mathbb{P}(\text{\ac{nlos}}^{(x)}) &=& 1-\mathbb{P}(\text{\ac{los}}^{(x)})  = 1-e^{-\lambda\eta^{(x)}\mathbb{E}[L]r}.
\end{eqnarray}
Note that the previous probabilities do not depend on the position of the point $x$ since $\eta^{(x)}$ does not depend on $x$ either.

\begin{figure}[tbp]
\centerline{\includegraphics[width=10cm]{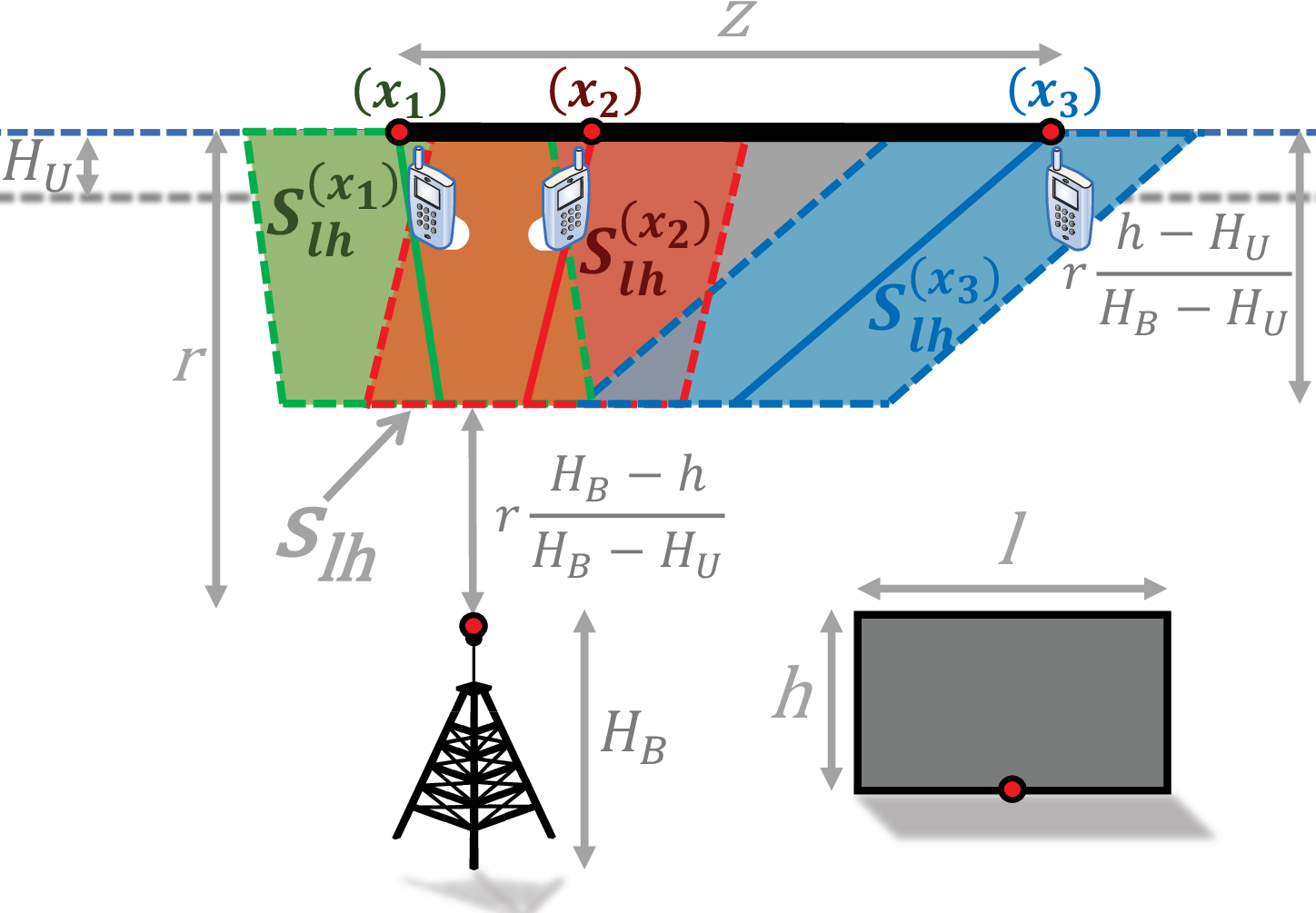}}
\caption{The shadowed area is the geometric locus of the centers of the blocking elements of length $l$ and height $h$ able to block, at least, one of the points in the black segment of length $z$.}
\label{fig:Blocking_region_general}
\end{figure}

Fig. \ref{fig:Blocking_region_general} shows in green, red, and blue colors the blocking regions for three points, namely $x_1, x_2, x_3$, within the depicted segment of length $z$. Let $K_{lh}$ denote the number of blocking elements with length $l$ and height $h$ within the whole shadow area in Fig. \ref{fig:Blocking_region_general}, and $S_{lh}$ the union of the blocking regions $S_{lh}^{(x)}$ of all the points in the segment.
For an entire segment to be in \ac{los}, all the points within this segment must be in \ac{los}. Therefore, the probability that a given segment is in \ac{los} is:
\begin{equation}
		\mathbb{P}(\text{a given segment is in \ac{los}})=\mathbb{P}(K=0)=e^{-\mathbb{E}[K]},\label{eq:prob_segment_los}
\end{equation}
where $K$ is the number of blocking objects (of any length and height) that block, at least, one of the points in the segment. $K$ is also a Poisson \ac{rv} whose mean value is derived as follows.

As the trajectory is parallel to the orientation of the blocking elements, the union of the blocking regions for all the segment points has an area equal to (this area is not equal to the sum of the areas of the individual regions since they overlap):
\begin{equation}\label{eq:A_S_lhtheta}
	A_{S_{lh}}= 
		\left\{ 
			\begin{array}{l}
         			\frac{r}{2}\left(z\left(1-\left(\frac{H_B-h}{H_B-H_U} \right)^2 \right)+2\frac{h-H_U}{H_B-H_U}l\right) \\ \hspace{4.3cm} \text{if} \hspace{0.3cm} H_U \leq h \leq H_B, \\
             			\frac{r}{2}\left(z + 2l\right) \hspace{3.7cm} \text{if} \hspace{0.3cm} H_B < h,
             		\end{array}
   		\right.
\end{equation}
where, as mentioned before, $z$ is the length of the segment. Accordingly, the mean value of $K$ is
\begin{equation}
    \mathbb{E}[K]=\lambda\int\int A_{S_{lh}} f_L(l) f_{H}(h) dl dh=\lambda \frac{r}{2} (\tilde{\eta} z+2\eta^{(x)}\mathbb{E}[L]),
\end{equation}
where the parameter $\tilde{\eta}$ is (see the Appendix for a proof)
\begin{equation}\label{eq:eta}
	\tilde{\eta}= 
		\left\{ 
			\begin{array}{l}
             			1 \hspace{5.2cm} \text{if} \hspace{0.2cm} H_B < H_{\min}, \\
             			\frac{H_{\max}-H_B}{H_{\max}-H_{\min}} + \frac{ (H_U^2-2H_BH_U)(H_B-H_{\min}) }{(H_{\max}-H_{\min})(H_B-H_U)^2} \\ \hspace{0.4cm}+ \frac{ \frac{2}{3}H_B^3-H_BH_{\min}^2+\frac{1}{3}H_{\min}^3 }{(H_{\max}-H_{\min})(H_B-H_U)^2} \hspace{0.2cm} \text{if} \hspace{0.2cm} H_{\min} \leq H_B \leq H_{\max}, \\
             			\frac{H_U^2-2H_BH_U}{(H_B-H_U)^2} - \frac{H_{\max}^2+H_{\max}H_{\min}+H_{\min}^2}{3(H_B-H_U)^2} \\ \hspace{0.4cm} + \frac{H_B(H_{\max}+H_{\min})}{(H_B-H_U)^2} \hspace{2.4cm} \text{if} \hspace{0.2cm} H_{\max} < H_B.
             		\end{array}
   		\right.
\end{equation}

Finally, the probability that the segment is in \ac{los} (\ref{eq:prob_segment_los}) is
\begin{equation}
	\mathbb{P}(\text{a given segment is in \ac{los}}) = e^{-\lambda \frac{r}{2} (\tilde{\eta} z+2\eta^{(x)}\mathbb{E}[L])},
\end{equation} 
which decreases exponentially with the density and average size of the blockages, the distance of the trajectory to the \ac{bs}, and the length of the segment.
\section{Statistical Characterization of LOS Intervals} \label{sec:Light} 
Our goal is to obtain the statistical distribution of the \ac{rv} $Z$, which stands for the length of \ac{los} intervals in a given trajectory at a distance $r$ from the \ac{bs}, that is, to obtain a closed-form expression of the \ac{pdf} $f_Z(z)$. To that end, we first compute the \ac{cdf} $F_Z(z)=1-\mathbb{P}(Z \geq z)$. 

A \ac{los} segment will contain, by definition, at least one point in \ac{los}. Based on that, without loss of generality, we consider a point $a$ that is in \ac{los}. Then, $\mathbb{P}(Z \geq z)$ is the probability that $a$ is within a \ac{los} segment of length, at least, $z$. Fig. \ref{fig:Overlapping_blocking_regions_general} 
shows an example with three valid segments. We want to compute the probability of the union of events, where each event corresponds to each segment being in \ac{los}. An exact expression for this probability is difficult to obtain because there is a continuous, infinite, and uncountable collection of segments with non-null overlap. However, we can state that:

\begin{figure}[tbp]
\centerline{\includegraphics[width=\columnwidth]{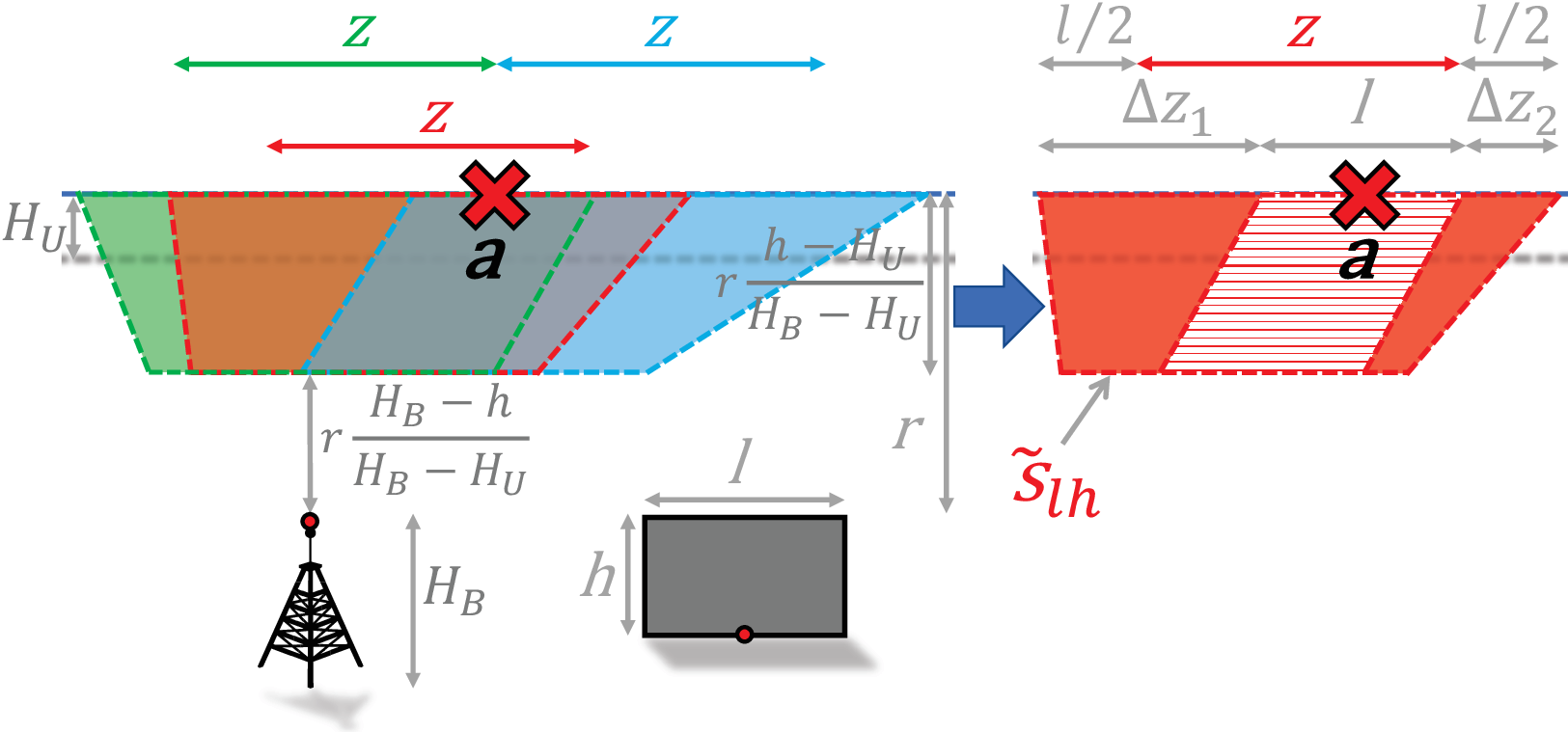}}
\caption{Overlapping blocking regions of three different segments of length $z$ that contain a given point $a$. The dark red region $\tilde{S}_{lh}$ (i.e., the union of the two dark trapezoids) is the geometric locus of the centers of the blocking elements that will block, at least, one point in the segment of length $z$ knowing that they do not block the point $a$.}
\label{fig:Overlapping_blocking_regions_general}
\end{figure}

\begin{equation} \label{eq:segment_in_light}
  \begin{aligned}[b]
      \mathbb{P}(Z \geq z)& =\mathbb{P}(\text{at least one segment of length } \geq z \\
      &\hspace{0.7 cm}\text{ containing }a\text{ is in \ac{los}}|a\text{ is in \ac{los}}) \\
      & \geq \max_{v \in \mathcal{V}} \mathbb{P}(v \text{ is in \ac{los}}|a\text{ is in \ac{los}}),
  \end{aligned} 
\end{equation}
where $\mathcal{V}$ is the set of segments of length $z$ containing $a$. The inequality in (\ref{eq:segment_in_light}) comes from the Fr\'echet bound \cite{Frechet} and the fact that the segment \ac{los} probability decreases with the segment length. Therefore, we have that:
\begin{equation}
  \begin{aligned}[b]
       F_Z(z) \leq 1-\max_{v \in \mathcal{V}} \mathbb{P}(v \text{ is in \ac{los}}|a\text{ is in \ac{los}}).
  \end{aligned} \label{eq:CDF}
\end{equation}

To compute (\ref{eq:CDF}), we need to obtain the probability that a segment $v \in \mathcal{V}$ (i.e., $v$ has length $z$ and contains the point $a$) is in \ac{los} when we know that the point $a$ is in \ac{los}. As in Section \ref{sec:prob_segm_los}, let us first consider blocking elements of length $l$ and height $h$. We can see from Fig. \ref{fig:Overlapping_blocking_regions_general} that $\mathbb{P}(v \text{ is in \ac{los}}|a\text{ is in \ac{los}})$ is equal to the probability that no points that represent the centers of the blocking elements fall within the dark red trapezoids since we know for sure that no points fall within the central lined parallelogram because point $a$ is in \ac{los}. As the trajectory is parallel to the orientation of the buildings, the area of the dark red region, denoted by $\tilde{S}_{lh}$ in Fig. \ref{fig:Overlapping_blocking_regions_general} (i.e. the union of the two dark trapezoids), is
\begin{equation}\label{eq:A_S_lhtheta_dark}
	A_{\tilde{S}_{lh}}= 
		\left\{ 
			\begin{array}{lcc}
             			\frac{r}{2}z\left(1-\left(\frac{H_B-h}{H_B-H_U} \right)^2 \right) &\text{if}& H_U \leq h \leq H_B, \\
             			\frac{r}{2}z &\text{if}& H_B < h.
             		\end{array}
   		\right.
\end{equation}

Note that the expression of $A_{\tilde{S}_{lh}}$ is the same for all the segments of length $z$ within a trajectory at distance $r$ from the \ac{bs}. Let $\tilde{K}$ be the number of blocking elements of any length and height that can block, at least, one point in a segment $v \in \mathcal{V}$, without blocking the point $a$. As $\tilde{K}$ is a Poisson \ac{rv} with mean $\mathbb{E}[\tilde{K}]=\lambda \int_l\int_hA_{\tilde{S}_{lh}}f_L(l) f_{H}(h) dl dh$, we have that
\begin{equation}
  \begin{aligned}[b]
      & \mathbb{P}(v \text{ is in \ac{los}}|a\text{ is in \ac{los}})= e^{-\mathbb{E}[\tilde{K}]}=e^{-\lambda \tilde{\eta} \frac{r}{2}z }, \text{ for } v \in \mathcal{V}.
  \end{aligned}
\end{equation}
Then, the \ac{cdf} of the length of the \ac{los} intervals fullfils
\begin{equation}
	F_Z(z)  =  \mathbb{P}(Z \leq z) = 1 - \mathbb{P}(Z \geq z) \leq 1 - e^{-\lambda \tilde{\eta} \frac{r}{2}z }.
\end{equation}
Assuming that the previous upper bound is tight, we can obtain an approximation of the actual \ac{pdf} deriving the bound:
\begin{eqnarray}\label{eq:pdf}
	f_Z(z) = \frac{dF_Z(z)}{dz} \approx \lambda \tilde{\eta} \frac{r}{2}e^{-\lambda \tilde{\eta} \frac{r}{2}z },
\end{eqnarray}
that is, $Z$ can be approximated as an exponential \ac{rv} parameterized by $\lambda$, $\tilde{\eta}$, and $r$.

\section {Average Length and Density of the \ac{los} and \ac{nlos} Intervals within a Trajectory} \label{sec:Average}
Let $S$ denote the \ac{rv} corresponding to the length of the blocked intervals. To calculate the \ac{pdf} of $S$, we should first compute the probability that there is at least one blocked segment of length greater than or equal to $s$ that contains a reference blocked point $a$:
\begin{equation}
  \begin{aligned}[b]
     \mathbb{P}(S \geq s)& =\mathbb{P}(\text{at least one segment of length } \geq s \\
          &\hspace{0.7 cm}\text{ containing }a\text{ is blocked}|a\text{ is in \ac{nlos}}).
  \end{aligned} \label{eq:Pshadow}
\end{equation}
The probability in (\ref{eq:Pshadow}) entails a continuous infinite and uncountable collection of segments, making its computation very difficult. We could focus instead on obtaining a pessimistic upper bound of (\ref{eq:Pshadow}). However, the use of Fréchet inequalities would lead to a lower bound of (\ref{eq:Pshadow}). Fortunately, it is possible to obtain a good approximation of the average number and average length of the blocked intervals in a given trajectory and evaluate its accuracy through numerical simulations. 

From (\ref{eq:pdf}), we can compute an approximation of the average length of the \ac{los} segments:
\begin{eqnarray}
	\mathbb{E}[Z] &=& \int_{0}^{\infty} \hspace{-0.2cm}z f_Z(z)dz \approx \int_{0}^{\infty} \hspace{-0.2cm} z \lambda \tilde{\eta} \frac{r}{2}e^{-\lambda \tilde{\eta} \frac{r}{2}z } dz =   \frac{2}{\lambda \tilde{\eta} r}.
\end{eqnarray}
We will check the validity of this approximation through numerical simulations in Section \ref{sec:Results}. Note that
\begin{align}
	&\lim_{r\to 0} \mathbb{E}[Z] = \infty, & \lim_{r\to \infty} \mathbb{E}[Z] = 0, \label{eq:limZ}
\end{align}
i.e., if the trajectory is close to the \ac{bs}, the blockage probability is much smaller and the average length of the \ac{los} intervals tends to infinity. The opposite happens when $r\to \infty$.

Accordingly, the average number of \ac{los} intervals within a trajectory of total length $d$ is:
\begin{equation}
	\mathbb{E}[N_z] = \frac{\mathbb{P}(\text{\ac{los}}^{(x)}) \cdot d}{\overline{Z}} \approx \lambda \tilde{\eta} r \frac{e^{-\lambda \eta^{(x)} \mathbb{E}[L]r}}{2}d, 
\end{equation}
Since after every \ac{los} interval, there is a NLOS interval, and viceversa; the average number of blocked intervals within a trajectory of length $d$ is $\mathbb{E}[N_s] = \mathbb{E}[N_z]$.

The average number of \ac{los} (and blocked) intervals depends on $r$. The maximum density of intervals is:
\begin{equation}
	\frac{\mathbb{E}[N_z]}{d}\Big|_{\max} = \frac{\mathbb{E}[N_s]}{d}\Big|_{\max} = \frac{\tilde{\eta}}{\eta^{(x)}}\frac{1}{2 \mathbb{E}[L]e},\label{eq:max_density}
\end{equation}
and is attained when $r=\frac{1}{\lambda \eta^{(x)}\mathbb{E}[L]}$.

Note that, when $r$ tends to 0, that is, the trajectory is very close to the \ac{bs}, the \ac{los} intervals tend to be much longer than the blocked ones and, therefore, their density (i.e., the average number of intervals in a trajectory divided by the length $d$ of the trajectory) is very low. On the other hand, when the trajectory is far from the \ac{bs}, blocked intervals are longer than the \ac{los} intervals, and the density becomes low again:
\begin{align}
	&\lim_{r\to 0} \hspace{-0.05cm}\frac{\mathbb{E}[N_z]}{d} \hspace{-0.05cm} = \hspace{-0.05cm} \lim_{r\to 0}\hspace{-0.05cm} \frac{\mathbb{E}[N_s]}{d} \hspace{-0.05cm}=\hspace{-0.05cm} 0, &\lim_{r\to \infty}\hspace{-0.05cm} \frac{\mathbb{E}[N_z]}{d}\hspace{-0.05cm} =\hspace{-0.05cm} \lim_{r\to \infty}\hspace{-0.05cm} \frac{\mathbb{E}[N_s]}{d}\hspace{-0.05cm} =\hspace{-0.05cm} 0.
\end{align}

We are also interested in estimating the average length of the blocked intervals. Since $\mathbb{E}[N_s]=\mathbb{E}[N_z]$, we have that:
\begin{equation}
	\mathbb{E}[S] \hspace{-0.05cm}=\hspace{-0.05cm} \frac{\mathbb{P}(\text{\ac{nlos}}^{(x)})}{\overline{N_s}}d \hspace{-0.05cm} = \hspace{-0.05cm} \frac{\mathbb{P}(\text{\ac{nlos}}^{(x)})}{\mathbb{P}(\text{\ac{los}}^{(x)})} \mathbb{E}[Z] 
	\hspace{-0.05cm} \approx \hspace{-0.05cm} \frac{2\hspace{-0.05cm} \left(\hspace{-0.05cm}e^{\lambda \eta^{(x)} \mathbb{E}[L]r}\hspace{-0.1cm}-\hspace{-0.1cm}1 \hspace{-0.05cm}\right)}{\lambda\tilde{\eta} r}\hspace{-0.1cm}, 
\end{equation}
with the following behavior, different from (\ref{eq:limZ}):
\begin{align}
	& \lim_{r\to 0} \mathbb{E}[S] = 2\frac{\eta^{(x)}}{\tilde{\eta}}\mathbb{E}[L], & \lim_{r\to \infty} \mathbb{E}[S] = \infty.
\end{align}

Another interesting aspect is the calculation of the point where the lengths of the \ac{nlos} and \ac{los} intervals are equal, in average, something that happens when $r = \ln2 \frac{1}{\lambda \eta^{(x)} \mathbb{E}[L]}$:
\begin{equation}
	\mathbb{E}[Z]\Big|_{r = \frac{\ln2}{\lambda \eta^{(x)} \mathbb{E}[L]}}  = \mathbb{E}[S]\Big|_{r = \frac{\ln2}{\lambda \eta^{(x)} \mathbb{E}[L]}}  = \frac{\eta^{(x)}}{\tilde{\eta}} \mathbb{E}[L] \frac{2}{\ln 2}.\label{eq:equal_aver_lengths}
\end{equation}

\section{Results} \label{sec:Results}
This section evaluates the tightness of the derived approximations. Firstly, we do simulations where the blocking elements are spawned according to a uniform spatial \ac{ppp} within an area delimited by the analyzed trajectory and its minimum distance $r$ to the \ac{bs}. We have considered $H_U= \SI{1.5}{m}$, $H_B= \SI{25}{m}$, $L\sim\mathcal{U}[L_{\min},L_{\max}]$ with $L_{\min} = \SI{10}{m}$ and $L_{\max}= \SI{30}{m}$, $H\sim\mathcal{U}[H_{\min},H_{\max}]$ with $H_{\min}= \SI{10}{m}$ and $H_{\max}= \SI{30}{m}$. Then, we have that $H_{\min} \leq H_B \leq H_{\max}$ which, as explained, will set the values for both $\eta^{(x)}$ and $\tilde{\eta}$. To study the impact of the blocker densities we consider two different values of $\lambda$. To set these values, we have taken as a reference the density of the buildings in one of the emblematic areas of Barcelona: the Eixample. In this area, the buildings are distributed in groups of four buildings every 1 or 1.24 hectares (i.e., the density is around to $3.22\cdot10^{-4}$ buildings$/m^2$).

\begin{figure}[tbp]
\centerline{\includegraphics[width=10cm]{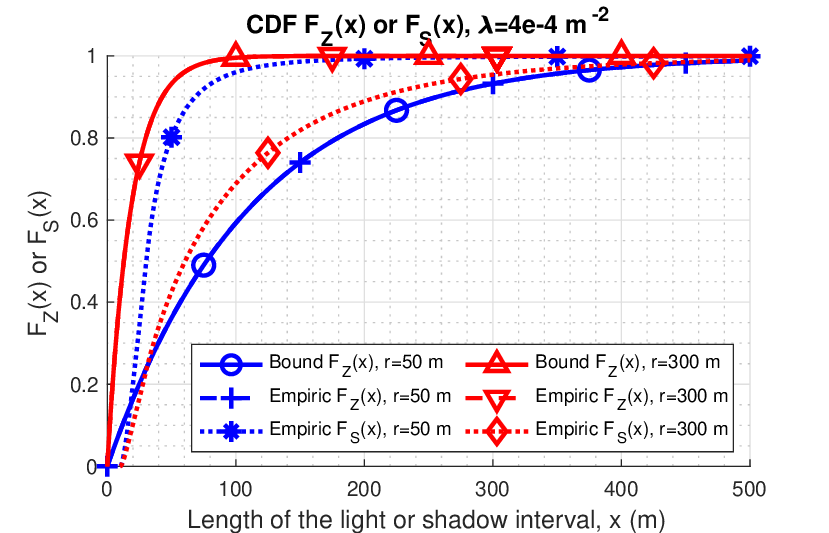}}
\caption{Comparison between the analytical upper bound and the empirical \ac{cdf}s.}
\label{fig:CDFs}
\vspace{-4mm}
\end{figure}

\begin{figure}[tbp]
\centerline{\includegraphics[clip, trim=1cm 0cm 1.3cm 0.3cm, width=12cm]{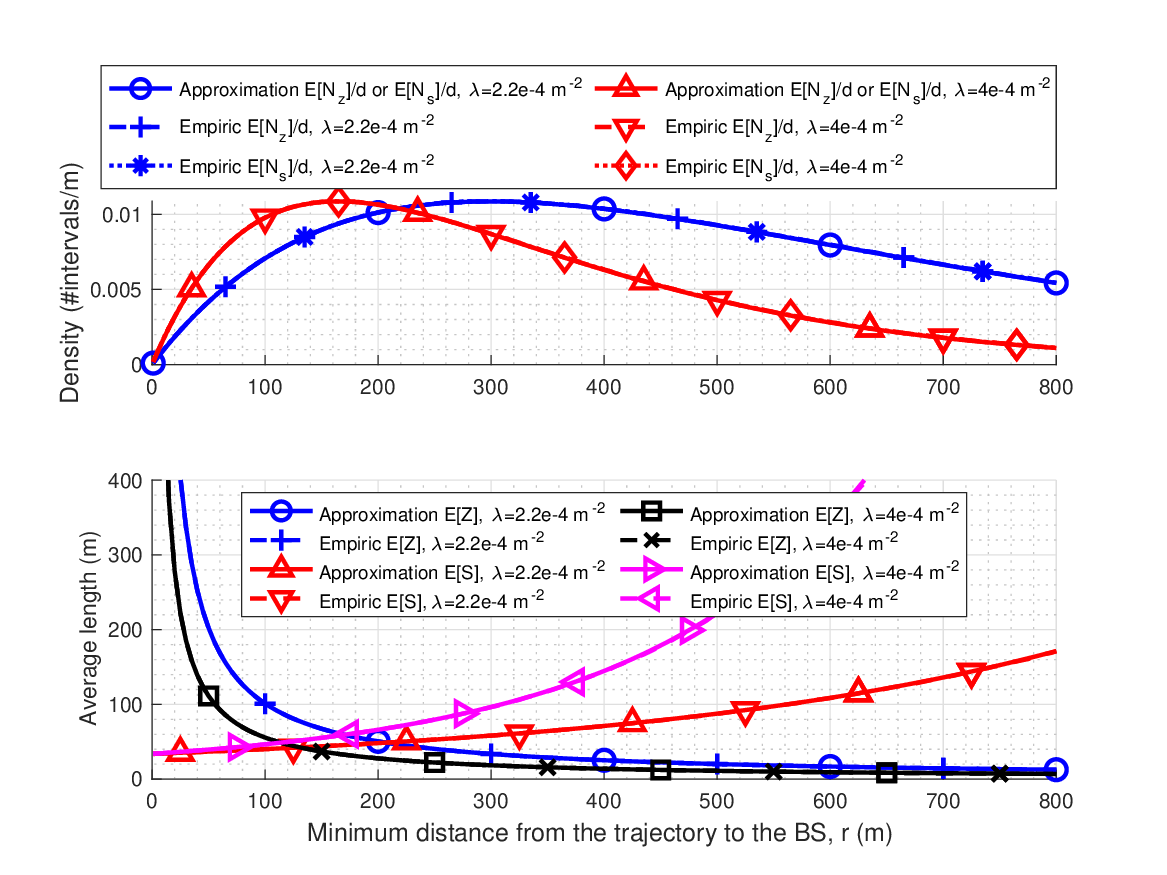}}
\caption{Density of \ac{los}/\ac{nlos} intervals and average lengths.}
\label{fig:Density_and_average_length}
\end{figure}
Fig. \ref{fig:CDFs} depicts the empirical CDFs of the lengths of the \ac{los} and \ac{nlos} intervals obtained from simulations, as well as the analytic upper bound obtained in Section \ref{sec:Light} for the \ac{los} \ac{cdf}. We see that the upper bound is very close to the numerical results. For a small $r$, the \ac{nlos} intervals are shorter, and the \ac{cdf} of the \ac{nlos} intervals length increases faster than the \ac{cdf} of the length of \ac{los} intervals. The opposite happens for high values of $r$. Fig. \ref{fig:Density_and_average_length} shows the dependence of the density and the average lengths of the intervals on $r$. Note that the theoretical approximations are very close to the simulation results. As predicted in Section \ref{sec:Average}, the maximum of the density of the LOS/NLOS intervals occurs when $r=\frac{1}{\lambda \eta^{(x)}\mathbb{E}[L]}$. Although the position of the maximum of the density of \ac{los} and \ac{nlos} intervals depends on $\lambda$, the value of the maximum does not (\ref{eq:max_density}). We also observe in Fig. \ref{fig:Density_and_average_length} that the average length of the blocked and LOS intervals become equal for a specific distance to the BS, $r = \frac{\ln2}{\lambda \eta^{(x)} \mathbb{E}[L]}$. While this distance depends on $\lambda$, the average length at this point does not (\ref{eq:equal_aver_lengths}).

\begin{figure}[tbp]
\centerline{\includegraphics[clip, trim=1.4cm 0.5cm 1.2cm 0.9cm, width=10cm]{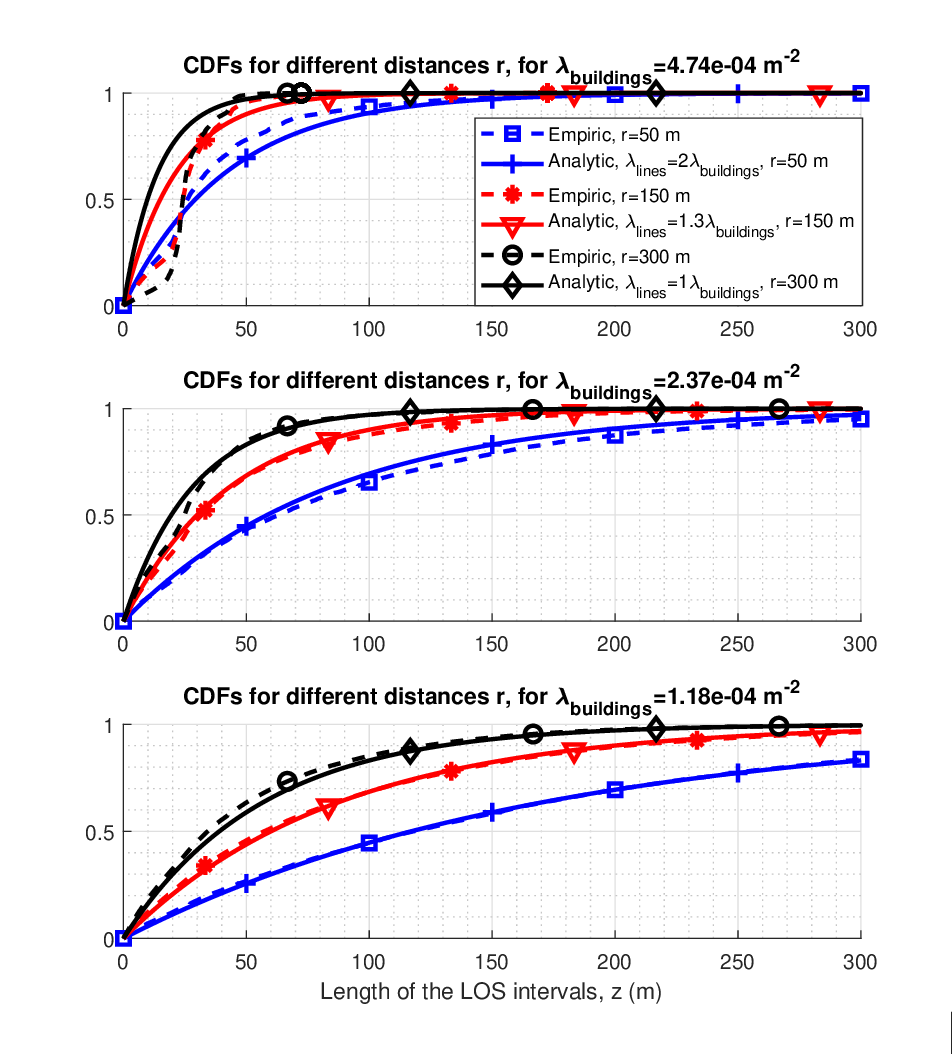}}
\caption{Empirical CDFs of the lengths of the LOS intervals for a real deployment in Chicago and analytical approximations.}
\label{fig:measurements_chicago}
\end{figure}

Finally, we have tested our analysis with the data in \cite{measurementsKulkarni} corresponding to a real layout of a Chicago area of size $\SI{1000}\times\SI{1000}{m}$. We have computed the \ac{cdf} of the lengths of the actual \ac{los} intervals in trajectories at different distances $r$ from the BS, located at random positions in the layout. To test our analytical expressions' fitness to different densities, we have worked with all the buildings, and also with a reduced number (by a factor two and four), leading the different densities of buildings $\lambda_{\text{buildings}}$. Fig. \ref{fig:measurements_chicago} shows the actual empirical CDFs obtained from the real-world layout and the analytical expressions. Note that the derivations in this paper assume that the blocking objects are lines. In order to generate the analytic CDFs in Fig. \ref{fig:measurements_chicago}, we assume that each building can be modelled as two parallel lines for low values of $r$ (such as $r=\SI{50}{\meter}$), so, $\lambda_{\text{lines}}=2\lambda_{\text{buildings}}$. This is because, for each building, the two sides causing most of the blocking, i.e., those parallel to the trajectory, produce 'shadows' without significant overlapping. Therefore, both of them should be considered. On the other hand, for distant trajectories, a high percentage of buildings will be far from the\ac{bs}. For a building far from the \ac{bs}, the individual 'shadows' generated by the two sides of the building that are parallel to the trajectory will have a very high percentage of overlapping that increases as $r$ increases. Therefore, each building can be modelled as just one line for high $r$ (such as $r=\SI{300}{\meter}$), so, $\lambda_{\text{lines}}=\lambda_{\text{buildings}}$. For intermediate values of $r$, the equivalent density of lines is in between (for example, for $r=\SI{150}{\meter}$, taking $\lambda_{\text{lines}}=1.3\lambda_{\text{buildings}}$ produces very adjusted results). 
Note that the analytic expressions match quite well the actual empirical CDFs. Only for a high value of $\lambda_{\text{buildings}}$, the shapes of the curves are not exactly equal. This discrepancy is because of the regularity of the real-world deployment, while the analytical expressions consider blocking elements following a spatial PPP. Also, for high distances to the BS and very high density, the length of the LOS intervals becomes limited by the width of the streets (in this deployment, around 30 meters approximately). In contrast, when the distance to the BS is lower, e.g., $r=\SI{50}{\meter}$, this limitation is not that important. Note that, for the highest densities and highest distances to the \ac{bs}, even though the empirical curves' shapes are not exactly equal to the analytic ones, they still accurately capture the values at which the empirical CDFs rise. 

\section{Conclusions}\label{sec:Conclusions}
In this paper, we have statistically characterized the lengths of the LOS and NLOS intervals within a trajectory in an urban scenario. Specifically, we have derived an upper bound of the \ac{cdf} of the length of the \ac{los} intervals that a user experiences when moving along a trajectory parallel to the orientation of the blocking elements, a likely scenario in real cities. 

We have found that the lengths of \ac{los} intervals approximately follows an exponential distribution with mean inversely proportional to the density of the blockages $\lambda$, the distance of the trajectory to the \ac{bs} $r$, and the $\tilde{\eta}$ parameter that captures the effect of the heights of the user, the buildings, and the \ac{bs}. When all the buildings are higher than the \ac{bs}, $\tilde{\eta}$ takes a maximum value one; whereas when the BS can be higher than some buildings, $\tilde{\eta}$ takes values below one since not all the buildings block the signal. Also, we have come up with simple closed-form approximations for the average lengths and density of the \ac{los}/\ac{nlos} intervals. These approximations are very close to the numerical results obtained with synthetic simulations and data from a layout of real buildings in Chicago.
 
By incorporating the users' speed, the results in this paper translate into information regarding the time a user will be in \ac{los}/\ac{nlos}. This information may be useful to make system design decisions regarding frame structure, \ac{ack} retransmission time, error correction codes, or handover management, among others. We leave as future work to consider macro-diversity with more than one \ac{bs}, the derivation of the outage value of the length of the \ac{nlos} intervals, and the analysis of the impact of distance on the received signal level.

\appendix

\subsection{Derivation of parameter $\eta^{(x)}$}
The value of $\mathbb{E}[K^{(x)}]= \int_l \int_h \lambda_{lh} A_{S_{lh}^{(x)}} $ in (\ref{eq:meanKx}), where the heights of the buildings follow a uniform distribution in the interval $[H_{\min},H_{\max}]$ and $A_{S_{lh}^{(x)}}$ is given by (\ref{eq:A_S_lhtheta_(x)}), is calculated for the following cases:\subsubsection{Case $H_B < H_{\min}$} in this case, all the buildings have a height $h>H_B$, thus,

\begin{equation}
\mathbb{E}[K^{(x)}] = \lambda r \int lf_L(l)dl \int f_H(h)dh = \lambda \mathbb{E}[L] r. \nonumber
\end{equation}

\subsubsection{Case $H_{\min} \leq H_B \leq H_{\max}$} in this case, some buildings have a height $h>H_B$ and some others $h\leq H_B$:
\begin{eqnarray}
	\mathbb{E}[K^{(x)}] &=& \frac{\lambda r\int l f_L(l)dl}{H_{\max}-H_{\min}} \left( \int_{H_{\min}}^{H_B} \hspace{-0.1cm}\frac{h-H_U}{H_B-H_U}dh+ \hspace{-0.1cm}\int_{H_B}^{H_{\max}} \hspace{-0.2cm}1 dh \right) \nonumber \\
	&\hspace{-1.8cm}=& \hspace{-0.8cm} \lambda \frac{2H_UH_{\min}\hspace{-0.1cm} -\hspace{-0.1cm} H_B^2 \hspace{-0.1cm}-\hspace{-0.1cm} H_{\min}^2\hspace{-0.1cm} + \hspace{-0.1cm}2H_BH_{\max} \hspace{-0.1cm} -\hspace{-0.1cm}2H_UH_{\max}}{2\left(H_B-H_U \right)\left(H_{\max}-H_{\min}\right)} \mathbb{E}[L] r.\nonumber
\end{eqnarray}

\subsubsection{Case $H_B > H_{\max}$} in this case, all the buildings have a height $h<H_B$, thus,
\begin{eqnarray}
	\mathbb{E}[K^{(x)}] &=& \frac{\lambda r}{H_{\max}-H_{\min}} \int l f_L(l)dl \int_{H_{\min}}^{H_{\max}} \frac{h-H_U}{H_B-H_U} dh \nonumber \\
&=& \lambda \frac{H_{\max}+H_{\min}-2H_U}{2\left(H_B-H_U \right)} \mathbb{E}[L] r.\nonumber
\end{eqnarray}

By identifying the terms in the expression $\mathbb{E}[K^{(x)}]= \lambda\eta^{(x)}\mathbb{E}[L]$, we obtain the values of $\eta^{(x)}$ shown in (\ref{eq:eta_x}).

\subsection{Derivation of parameter $\tilde{\eta}$}
Note that the area $A_{S_{lh}}$ in (\ref{eq:A_S_lhtheta}) can be expressed as $A_{S_{lh}}=A_{S_{lh}^{(x)}}+A_{\tilde{S}_{lh}}$, where $A_{\tilde{S}_{lh}}$ is defined in (\ref{eq:A_S_lhtheta_dark}). Consequently, using the previous result in the Appendix, we have that $\mathbb{E}[K]= \int_l \int_h \lambda_{lh} (A_{S_{lh}^{(x)}}+A_{\tilde{S}_{lh}}) =\mathbb{E}[K^{(x)}] + \mathbb{E}[\tilde{K}]= \lambda\eta^{(x)}\mathbb{E}[L]+\int_l \int_h \lambda_{lh} A_{\tilde{S}_{lh}}$ with $\mathbb{E}[\tilde{K}] \triangleq \int_l \int_h \lambda_{lh} A_{\tilde{S}_{lh}}$. We now focus on the computation of $\mathbb{E}[\tilde{K}]$:

\subsubsection{Case $H_B < H_{\min}$}
\begin{equation}
    \mathbb{E}[\tilde{K}] = \lambda\frac{r}{2}z \int \int f_H(h)f_L(l)dhdl=\lambda\frac{r}{2}z.\nonumber
\end{equation}

\subsubsection{Case $H_{\min} \leq H_B \leq H_{\max}$}
\begin{eqnarray}
    \mathbb{E}[\tilde{K}]  & = &  \frac{\lambda r z\int f_L(l)dl}{2(H_{\max}-H_{\min})} \times\nonumber  \\ && \left( \int_{H_{\min}}^{H_B} 1-\left(\frac{H_B-h}{H_B-H_U} \right)^2 dh + \int_{H_B}^{H_{\max}} 1 dh \right). \nonumber
\end{eqnarray}

\subsubsection{Case $H_B > H_{\max}$}
\begin{equation}
    \mathbb{E}[\tilde{K}] = \frac{\lambda r z\int f_L(l)dl}{2(H_{\max}-H_{\min})} \int_{H_{\min}}^{H_{\max}}    1-\left(\frac{H_B-h}{H_B-H_U} \right)^2 dh.\nonumber
\end{equation}

By developing the previous polynomial integrals and identifying the terms in $\mathbb{E}[\tilde{K}] =\lambda \frac{r}{2} \tilde{\eta} z$, we obtain the expressions for $\tilde{\eta}$ detailed in (\ref{eq:eta}).

\bibliography{blocking_references}
\bibliographystyle{IEEEtran} 

\end{document}